\documentclass{article}
\newcommand\rcsinfo{ \$Source: /home/axel/paper/frequency/frequency.tex,v $ $ --
\$Revision: 1.12.3.16 $ $ -- \$Date: 2004/11/30 13:19:47 $ $ }

\usepackage{amsmath}
\usepackage{natbib}
\usepackage[dvips]{epsfig}
\newcommand{\url}[1]{\texttt{#1}}
\newcommand{\erf}{\mathop{\textrm{erf}}}
\begin{document}

\author{A.~G.~Rossberg\footnote{\url{<Axel@Rossberg.net>}}\\
  \textit{Freiburger Zentrum f{\"u}r Datenanalyse und Modellbildung (FDM),}\\
\textit{Eckerstr. 1, 79104 Freiburg, Germany}}
\date{\small Published in  \textit{International Journal of Bifurcation and
  Chaos}, \textbf{14}(6), 2115-2123 (2004)}
\title{On the limits of spectral methods for frequency estimation}

\maketitle

An algorithm is presented which generates pairs of oscillatory random
time series which have identical periodograms but differ in the number
of oscillations.  This result indicate the intrinsic limitations of
spectral methods when it comes to the task of measuring frequencies.
Other examples, one from medicine and one from bifurcation theory, are
given, which also exhibit these limitations of spectral methods.  For
two methods of spectral estimation it is verified that the particular
way end points are treated, which is specific to each method, is, for
long enough time series, not relevant for the main result.

\newpage

\section{Introduction}
\label{sec:intro}

Nearly everywhere in science we encounter the problem of determining
the frequency of oscillation of some autonomously oscillating system
from which a signal has been measured over a finite time period.
Most commonly, spectral methods are used: From the squared modulus of
the Fourier transform of the signal (the periodogram) its expectation
value (the power spectrum) is estimated, and the frequency is
determined by some characterization of the position of a peak in the
power spectrum \citep[see,
e.g.,][]{mo88:_compar_doppl,muzi93:_quant_heart,timmer96:_quant_tremor,mendez98:_differ_frequen_kiloh_qpos,
  godano99:_sourc_strom_vulcan_islan_isole_eolie_italy,
  korhonen02:_estim_cardiovasc, slavic02:_measur_turbo_wheel}.
%1
For linear systems driven by white noise it is easily seen that the
periodogram or, equivalently, the empirical autocorrelation function
do, in fact, contain all relevant information
\citep{brockwell91:_time_series,priestley81:_spect}.

For nonlinear oscillators, this is generally not the case.  As was
shown recently \citep{rossberg02:_power_not_frequency}, the power
spectrum of a noisy or chaotic (i.e., not perfectly periodic)
nonlinear oscillator which oscillates with a given average frequency ,
i.e. the number of cycles per unit time, can be completely arbitrary.
In particular, the power spectrum is unrelated to the frequency.  The
number of oscillation cycles is here defined by the number of
transition of the trajectory of the signal in delay space through a
Poincar{\'e} section.  Since this corresponds to a purely topological
relation between the trajectory and the Poincar{\'e} section, the
corresponding measure of frequency was named \emph{topological
  frequency} $\omega_{\text{top}}$.  Of course, intuition and
experience suggest that, for typical experimental time series, the
result of using some standard method to determine the oscillation
frequency from the power spectrum would not differ too much from the
topological frequency.  But from the power spectrum alone no upper
limit for this difference can be derived.  Nor seems any other
efficient method to be known which would allow to determine an upper
limit for this difference.

After giving some examples for oscillations which exhibit this
unrelatedness of topological frequency and power spectrum in
Section~\ref{sec:examples}, a result is presented which is in some
sense stronger than the result cited above.  A precise determination
of frequency is not only impossible from the power spectrum of a time
series, but also from the periodogram.  Since the power spectrum is,
in one way or another, estimated from the periodogram by an averaging
procedure, the estimated power spectrum contains less information than
the periodogram.  One might conjecture that from the additional
information contained in the periodogram, such as higher order
correlations between the amplitudes of individual Fourier modes,
frequency information can be extracted.  As is shown in
Section~\ref{sec:periodogram}, this is not the case: Examples for time
series are given which obviously have different topological
frequencies, but, by construction, \emph{identical} periodograms.  The
use of the periodograms of the raw data is justified for spectral
estimation only if end effects can be neglected, for example, when the
time series are long enough.  In Section~\ref{sec:estimators} it is
shown that, in fact, for long enough time series the result holds
valid also when end effects are explicitly taken account by spectral
estimators.

\section{Oscillation frequencies and power spectra of some nonlinear 
  oscillators}
\label{sec:examples}

\begin{figure}[h]
  \centering
  \epsfig{file=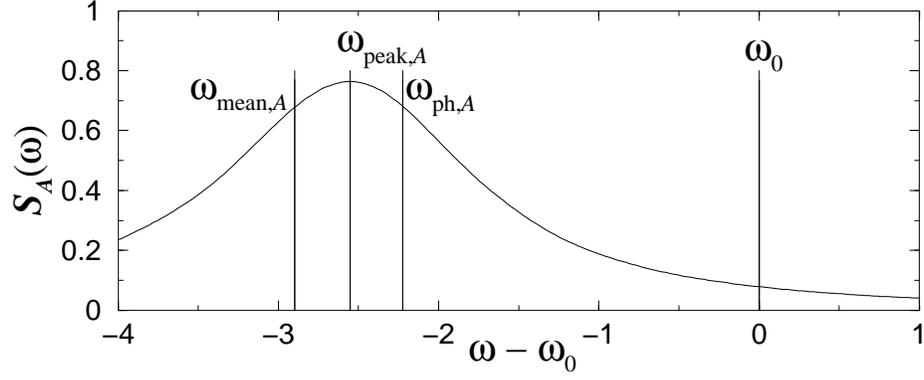,width=\columnwidth}
  \caption{The power spectrum $S_A(\omega)$ of $A(t)$ given by
    Eqs.\ (\ref{Hopf_normal_form},\ref{white_noise}) with $\epsilon=2$
    and $g_i=1$, obtained from a numerical simulation, and
    $\omega_{\text{peak},A}$ [see
    \cite{seybold74:_theor_detun_singl_mode_laser_near_thres} for
    analytic results], compared to the mean frequency
    $\omega_{\text{mean},A} = \omega_0 -
    g_i\left<|A|^4\!\right>/\left<|A|^2\!\right>$ [$\left< |A|^{2n}
      \!\right>=2^n\,\mathcal{N}^{-1}\,d^n\!\mathcal{N}/d\epsilon^n$,
    where $\mathcal{N}:=\pi^{1/2}\exp(\epsilon^2/4)
    (1+\erf(\epsilon/2))$, see
    \cite{risken89:_fokker_planc_equat__chap12}] and the phase
    frequency $\omega_{\text{ph},A}=\omega_0- g_i \left< |A|^2
      \!\right>$, defined by Eqs.\ 
    (\ref{define_f_phase},\ref{define_f_mean}), and the linear
    frequency $\omega_0$.}
  \label{fig:peak_vs_phase_f}
\end{figure}

As a first example, consider the noisy, weakly-nonlinear oscillator
described by a complex amplitude $A(t)$ with dynamics given by the
noisy Landau-Stuart equation
\citep{risken89:_fokker_planc_equat__chap12}
\begin{align}
  \label{Hopf_normal_form}
  \dot A = (\epsilon +i \omega_0) A - (1+i g_i)|A|^2 A + \eta(t),
\end{align}
where $\epsilon$, $\omega_0$, and $g_i$ are real and $\eta(t)$
denotes complex, white noise with correlations
\begin{align}
  \label{white_noise}
    \left< \eta(t)\eta(t') \right>=0,\quad
    \left< \eta(t)\eta(t^\prime)^*
    \right>=4 \delta(t-t^\prime)
\end{align}
[$^*$ $\equiv$ complex conjugation, $\left<\cdot \right>$ $\equiv$
expectation value].  In a certain sense
\citep{arnold98:_random_dynam_system}, this system universally
describes noisy oscillations in the vicinity of a Hopf bifurcation.
% \cm{$g D T^2\ll1$}.  
%
In general (i.e., with $g_i\ne 0$) the linear frequency $\omega_0$,
the \emph{spectral peak frequency} $\omega_{\text{peak}}$, i.e. the
position of the maximum of the power spectrum $S_A(\omega)$, the
average frequency or \emph{phase frequency}
\begin{align}
  \label{define_f_phase}
  \omega_{\text{ph},A}:= \left< \omega_i \right>,\quad
  \omega_i:=\mathrm{Im}\left\{\smash[t]{\dot A}/A\right\},
\end{align}
and
the \emph{mean frequency} of $A(t)$
\begin{align}
  \label{define_f_mean}
  \omega_{\text{mean},A}:=
  \frac{  
    \left<
      \omega_i |A|^2
    \right>
    }{\left<
      |A|^2
    \right>
    }
  = \frac{
    \mathrm{Im}
      \left<
        \smash{\dot A} A^*
      \right>
      }{
    \left<
      |A|^2
    \right>}
  =\frac{\int \omega S_{
      A}(\omega)d\omega}{\int S_{A}(\omega)d\omega},
\end{align}
are all different; see Fig.~\ref{fig:peak_vs_phase_f}.
[Definitions~(\ref{define_f_phase},\ref{define_f_mean}) are sometimes
restricted to \emph{analytic signals}, characterized by
$S_A(\omega)=0$ for $\omega \le 0$, which are in one-to-one
correspondence with the zero-mean, real-valued signals
$\mathrm{Re}\{A(t)\}$.
See~\cite{boashash92:_estim_inter_instan_frequen} for the history.]

The phase frequency measures the average number of circulations around
the point $A=0$ in phase space per unit time (decompose
$A(t)=a(t)e^{i\phi(t)}$ to see this).  Thus, with the natural
assumption that a circulation around the origin of phase space
corresponds to a cycle of oscillation -- which is justified by symmetry
considerations -- the phase frequency measures cycles per unit time,
i.e., the oscillation frequency.  Although this concept of an
oscillation cycle is slightly different from that used for the
definition of topological frequency, both frequency measures often
give identical results, in particular when the oscillation amplitude
does not come too close to zero.  For a discussion of the effect of
filtering on the phase frequency, see
\cite{rossberg02:_power_not_frequency}.  In contrast to the phase
frequency, the spectral peak frequency $\omega_{\text{peak},A}$ and
the mean frequency $\omega_{\text{mean},A}$ are frequency measures
which can be obtained directly from the power spectrum.  As can be
seen from Eqs.~(\ref{define_f_phase}) and (\ref{define_f_mean}),
$\omega_{\text{ph},A}=\omega_{\text{mean},A}$ if and only if
$\omega_i$ and $|A|^2$ are uncorrelated.  This is always the case for
linear, noise-driven oscillators, but for general nonlinear
oscillators it is not.

When it is known that the dynamics of $A(t)$ is of the form of
Eqs.~(\ref{Hopf_normal_form}) and (\ref{white_noise}) and the power
spectrum $S_A(\omega)$ of $A(t)$ is precisely determined, then it is
of course possible to obtain the free parameters of
Eq.~(\ref{Hopf_normal_form}) from a fitting procedure and to use these
to determine $\omega_{\text{ph},A}$.  But none of these assumptions is
typically satisfied in practice.  Equations~(\ref{Hopf_normal_form})
and (\ref{white_noise}) are exact only in a particular limit, and
$S_A(\omega)$ is usually biased by some nontrivial transfer function
inherent in the measurement process.  Then the information about
$\omega_{\text{ph},A}$ cannot be extracted from the power spectrum
anymore.  The situation becomes aggravated by the fact that, as was
shown by \cite{seybold74:_theor_detun_singl_mode_laser_near_thres},
$S_A(\omega)$ is always well approximated by a Lorenzian, uniformly
for all $\epsilon$.

As a second example, an experimental time series from the hand tremor
of a subject diagnosed with Parkinson's disease shall be examined.
The data was recorded by an accelerometer attached to the hand.  It
consists of $10240$ samples recorded at a rate of $300\,\mathrm{Hz}$.
The data is available on the
internet\footnote{\url{http://webber.physik.uni-freiburg.de/$\sim$jeti/tremordaten/park/0308-2-ali.dat}}.

\begin{figure}[h]
  \centering
  \epsfig{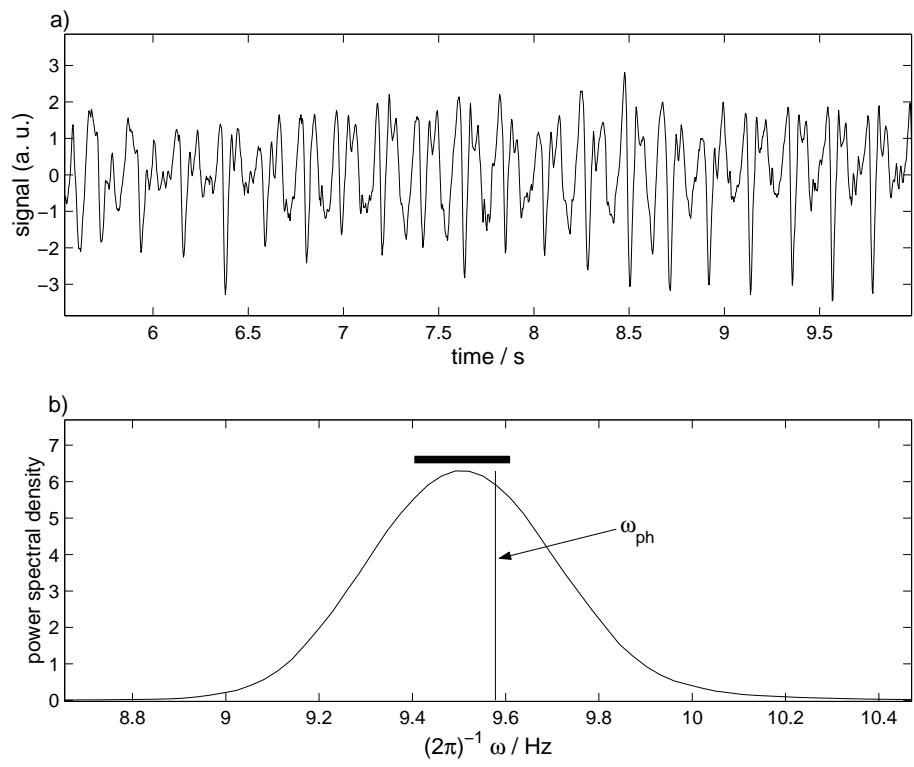}
  \caption{Power spectrum and oscillation frequency (b) of a human hand-tremor time series (a).   In part (b), the vertical line corresponds
    to the phase frequency of the time series, the horizontal bar
    shows the 95\% confidence interval for the spectral peak
    frequency.}
  \label{fig:tremor}
\end{figure}

Figure~\ref{fig:tremor}a shows part of the time series.  In order to
obtain an estimate of the power spectrum (Fig.~\ref{fig:tremor}b), the
periodogram of the (tapered) time series was smoothed using a window
of width $h=10$ with weights $w(i)=\max(0,1-i^2/h^2)$.  Its peak
frequency is at $\omega_{\text{peak}}/2\pi=9.49\,\mathrm{Hz}$,
corresponding to 324 oscillations over the time series.  In order to
determine the phase frequency, the (band-pass filtered) analytic
signal corresponding to the time series is obtained by convoluting it
with the Morlet wavelet
\begin{align}
  \label{morlet}
  w(t)=\frac{1}{\pi} \exp\left(i \omega_0 t - \frac{1}{2}
    (\Delta\omega t)^2\right)
\end{align}
with $\omega_0=\omega_{\text{peak}}$ and
$\Delta\omega=\omega_{\text{peak}}/6$.  The resulting phase frequency
$\omega_{ph}=9.58\,\mathrm{Hz}$ (Fig.~\ref{fig:tremor}b) is robust
under large variations of the filter width $\Delta\omega$ and mid
frequency $\omega_0$.  It is also identical to the topological
frequency obtained by band-pass filtering the time series by a
convolution with $\mathrm{Re}\{w(t)\}$ and a successive 2D delay
embedding, e.g., with delay $\tau=(23/300)\,\mathrm{s}$.
Unfortunately, the investigated time series is too short to yield the
observed difference between peak frequency and topological frequency
significant in a statistical sense, as a careful statistical analysis
using the method of \cite{timmer97:_confid} shows: The $95\%$
confidence interval for the peak frequency extends from
$9.40\,\mathrm{Hz}$ to $9.61\,\mathrm{Hz}$.  It would be desirable to
repeat the analysis on longer samples.

It should be emphasized that the point of
\cite{rossberg02:_power_not_frequency} is that the topological or
phase frequency cannot be derived from the power spectrum \emph{as a
  whole}.  Determination of the spectral peak frequency is only the
most obvious attempt to do so.  The failure of this particular
approach has been noticed before, see, e.g.,
\cite{pikovsky97:_phase_sync}.

\section{Does the periodogram help?}
\label{sec:periodogram}

Is it possible to identify a relative shift between spectral peak
frequency and topological or phase frequency by using a different
method to evaluate the periodogram?  As mentioned before, the answer
to be given here is negative: there is \emph{no} general way to tell
from the periodograms (or the empirical autocorrelation functions) of
time series that they oscillate with different frequencies.

For simplicity, only real valued time series $x(t)$ shall be
considered here for which the number of oscillations in a given time
interval of length $T$ can be determined by a simple 2D delay
embedding.  If there is a delay $\Delta t$, such that the trajectory
of the delay vector $(x(t),x(t-\Delta t))$ encircles the origin of the
coordinate system at finite distance, the number of circulations $m$
gives the number of oscillations in the time series.  The resulting
frequency is then $2\pi m/T$.  This is the definition of topological
frequency restricted to 2D embedding.  It is very similar to the one that
can be obtained from of the definition of instantaneous phase
$\phi(t)$ by \citet{pikovsky97:_phase_sync} when defining $\omega$ as the
temporal average of $d\phi/dt$.  However, their definition
is based on the actual phase space of a dynamical system and not on
its approximate reconstruction from a time series by delay embedding.
This distinction becomes particularly relevant when random
time series, as they are modeled by the method described here, are
evaluated.

We shall now outline an iterative algorithm that randomly generates
pairs of time series $x_1(t)$, $x_2(t)$ of the type described above.
The two time series generated by this algorithm have identical
periodograms but differ in the number of oscillation cycles.  (The
algorithm can also be considered as a source of random time series
$X_1(t)$, $X_2(t)$ which have identical power spectra but different
expectation values for the oscillation frequencies -- in fact, the
frequencies are fixed).
The algorithm uses some ideas from an algorithm of 
\citet{schreiber96:_improv_surrog_data_nonlin_tests} that generates
time series with periodograms nearly identical and distributions of
values identical to a given times series.

The main part of our algorithm generates two complex-valued random
time series $y_1(t)$, $y_2(t)$ with $N$ samples at $t=0,\ldots,N-1$.
When continued periodically [$y_l(t)=y_l(t+N)$], the distribution of
the output is invariant under translation in time.  The periodograms
of the two time series are identical.  The moduli $|y_l(t)|$ follow a
predetermined narrow distribution with values near one, i.e., the
values of $y_l(t)$ lie near the unit circle in the complex plane.
$y_1(t)$ and $y_2(t)$ encircle the origin of the complex plane $k_1$
and $k_2$ ($k_1 \ne k_2$) times respectively.  Using these time
series, the oscillatory, real-valued time series $x_1(t)$, $x_2(t)$
($x_l(t)=\mathrm{Re}\{\exp(i\omega_0 t) y_l(t)$) with identical
periodograms are constructed.
% , are then obtained as
% $X^\prime_l(t)=\mathrm{Re}\{\exp(i\omega_1 t) Y_l(t)\}$, with some not
% too small $\omega_1$.  After applying a small correction to the
% $X^\prime_l(t)$, the desired pair of real-valued time series $X_l(t)$
% with identical periodograms is obtained.

In order to obtain the final time series $y_l(t)$, the following two
basic steps are performed iterative, starting with two time series
which encircle the origin of the complex plane $k_1$ respectively
$k_2$ ($k_1 \ne k_2$) times, but do not yet have identical
periodograms: In Fourier space, the magnitudes of the Fourier modes of
the two time series are adjusted, such that the magnitudes of the
Fourier modes at the same frequencies approach each other, while the
phases of the Fourier modes are kept constant (Steps~\ref{startloop}
and~\ref{retransform} in the detailed description below).  Then, back
in physical ``space'', i.e., using the actual time series, a
nonlinear, monotonously increasing transformation of the magnitudes of
the values for each time is applied in such a way that the desired
distribution of the complex time series near the unit circle (see
above) is enforced (Step~\ref{endloop} below).  If the phases were
kept constant in this second basic step, the algorithm would (as
numerical experiments show) typically converge to a trivial solution
with $y_1(t)=y_2(t)$.  Obviously, the number of circulations of the
origin of the complex plane would then have changed with respect to
the initial condition -- at least for one of the two time series.  In
order to avoid these phase slips, the difference in phase angle
between successive values is, for both time series, enforced to lie
within a certain, small rage, by adjusting, at each iteration, also
the phase angles
(Steps~\ref{initial_perm},\ref{start_other}-\ref{newphases}).  The
details of these manipulations of the phases, as well as the precise
choice of the initial time series (Steps~\ref{initialize}
and~\ref{noisify}), contain random elements, as will be explained
below.  Experience shows that convergence of the algorithm is,
although not guaranteed, with appropriate choices of the tuning
parameters of the algorithm (see below), highly probable.

\begin{figure}[h]
  \centering
%   a)\hspace*{\fill}\ \\
   \epsfig{file=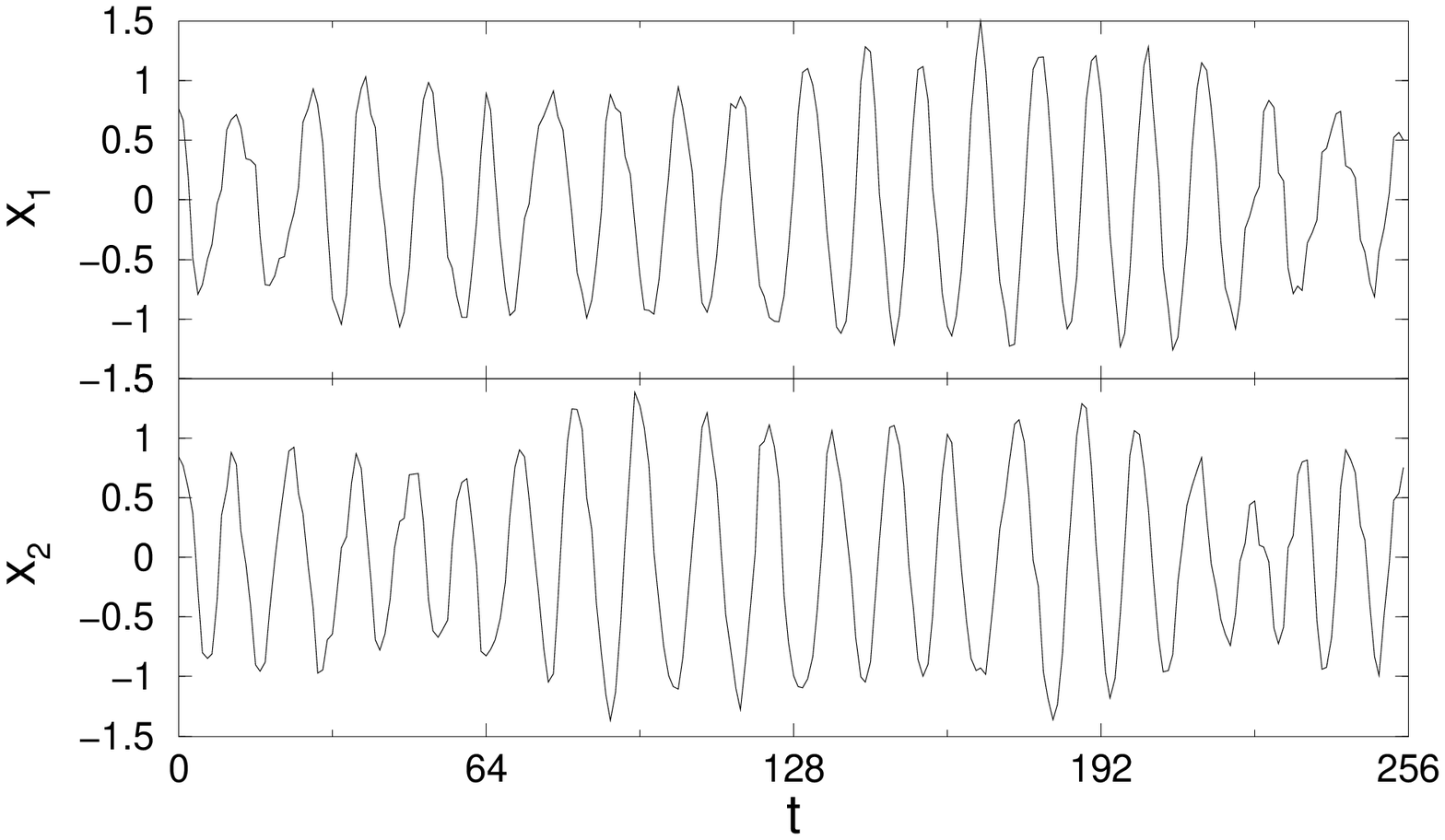,width=\columnwidth}\\
%   \epsfig{file=fig3a.eps,width=\columnwidth}\\
%   b)\hspace*{\fill}\ \\
   \epsfig{file=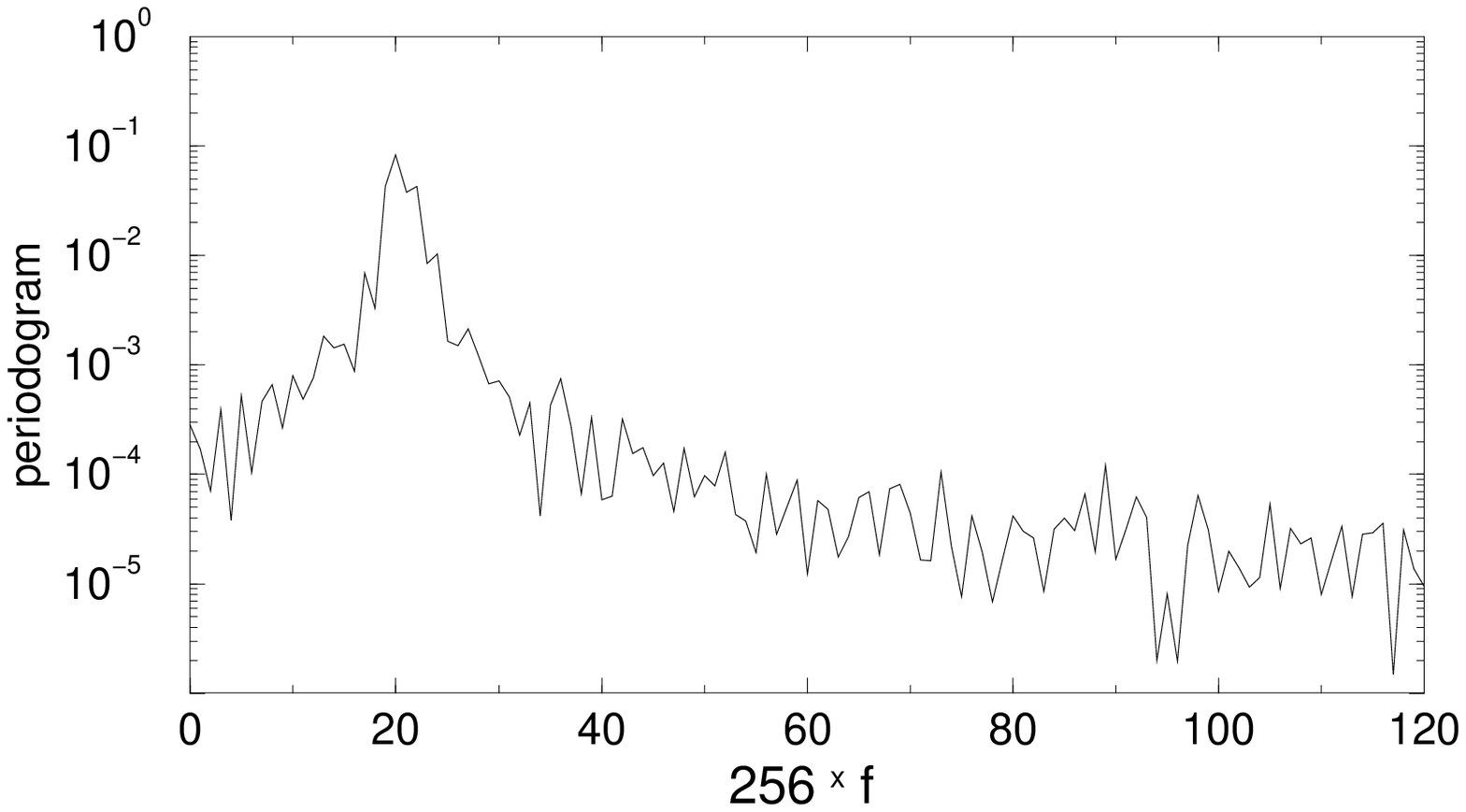,width=\columnwidth}\\
  \caption{(a)  A pair of random time series
    $x_1(t)$, $x_2(t)$ (\textit{Pair~A}) with frequencies $f=20/256$
    and $f=21/256$ respectively.  It was generated using the algorithm
    described in this report with the proposed parameters.  (b) The
    periodogram, which is identical for both time series.}
  \label{fig:random}
\end{figure}

\begin{figure}[h]
  \centering
  \epsfig{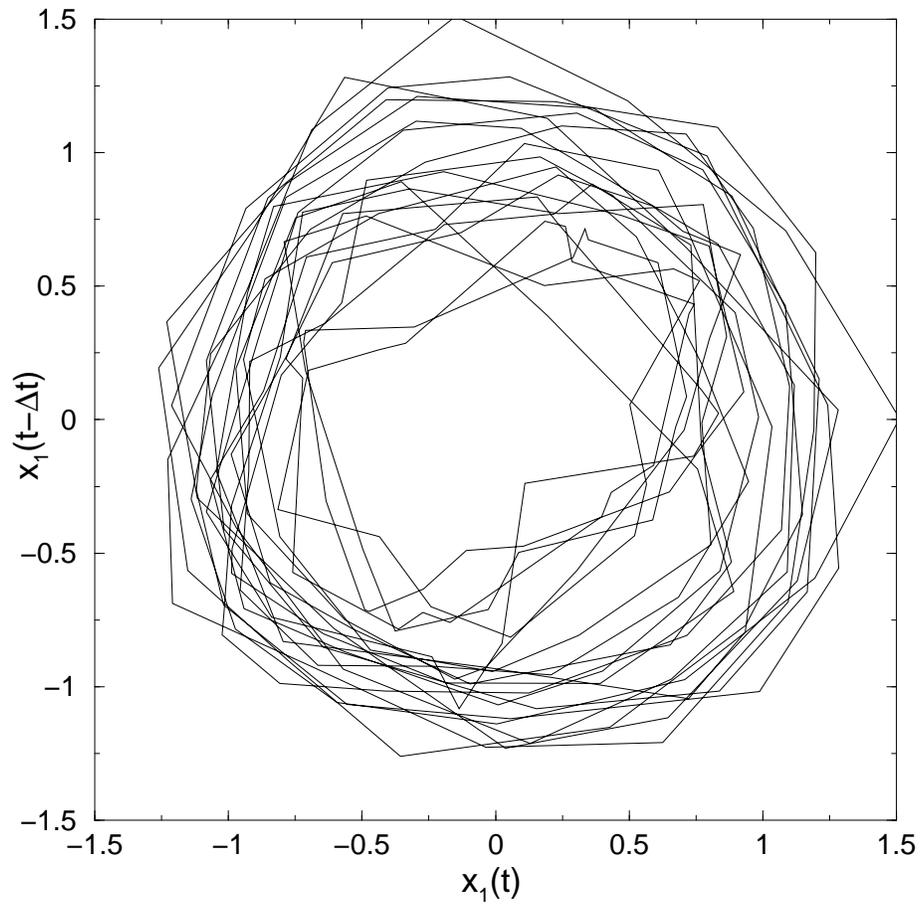}\\
  \caption{The trajectory of $x_1(t)$ in delay space with $\Delta
    t=3$.  By the fact that it is winding $20$ times around the origin
    at a finite distance, the frequency is sharply defined.}
  \label{fig:phasespace}
\end{figure}

This is a step-by-step description of this algorithm.  We give the
values of tuning parameters as they were used to generate the sample
\textit{Pair~A} shown in Fig.~(\ref{fig:random}), which consists of
two time-series with $N=256$ points, in parenthesis.
\begin{enumerate}
\item \label{initial_perm}Generate two random permutations
  $\Pi_l:(0\ldots(N-1))\to(0\ldots(N-1))$ ($l=1,2$) of length $N$,
  with equal probability $1/N!$ for each permutation, to be used
  later.
\item \label{initialize}Initialize the series $y_1(t)$ and $y_2(t)$ as
  $y_l(t)=\exp(2 \pi i k_l t/N)$ at $t=0,\ldots,N-1$ with small
  integer $k_l$ (e.g., $k_1=0$ and $k_2=1$).
\item \label{noisify}In order to excite also the remaining Fourier
  modes, add some noise to $y_1(t)$ and $y_2(t)$ (e.g., complex
  Gaussian with variance $2\times10^{-4}$).

  Now the main loop (steps \ref{startloop}-\ref{endloop}) starts.
\item \label{startloop} Calculate the discrete Fourier transforms
  $\tilde y_l(\omega)$ of the time series $y_l(t)$.  Increase or
  decrease the values of $|\tilde y_l(\omega)|$ such that the distance
  between $|\tilde y_1(\omega)|$ and $|\tilde y_2(\omega)|$ decreases
  by a fixed percentage (e.g., $20\%$) while keeping $\arg( \tilde
  y_l(\omega))$ and $|\tilde y_1(\omega)|+|\tilde y_2(\omega)|$ fixed.
  In this step the two periodograms are forced to approach each other.
  Notice, however, that the manipulations in steps
  \ref{start_other}-\ref{endloop} may have an adversary effect.  In
  order to have the algorithm as a whole converge, some balancing of
  the tuning parameters is therefore required.
\item \label{retransform} Obtain the inverse Fourier transforms of the
  results and assign them to $y_1(t)$ and $y_2(t)$.
\item \label{start_other} Calculate the increments of $\arg(y_l(t))$
  between successive samples.  Apply the permutations $\Pi_l$ to the
  sequences of phase increments (see below for an explanation).
\item \label{diffusion} For phase increments of absolute value larger
  than some threshold (e.g., 0.4) distribute part of the increment
  (e.g., $1\%$) symmetrically over the preceeding and succeeding
  increments such that the sum of the three remains unchanged.  Do
  this step with parallel update for all increments. 
% \item \label{diffusion} If some phase increments exceed a certain
%   threshold (e.g., 0.4), decrease their values (e.g., by $1\%$) and
%   symmetrically increase the values of the preceding and succeeding
%   increments, such that the sum remains unchanged.  Increase the
%   values of phase jumps correspondingly, where their negative value
%   exceeds the threshold.
\item \label{diffusion_loop} Repeat step \ref{diffusion} until no
  phase increment exceeds the threshold.  
  
\item \label{newphases} Obtain the inverse permutations $\Pi_l^{-1}$
  of these phase increments, and sum them up to obtain two new time
  series for the phases $\arg(y_l(t))$.
\item \label{endloop} For each series, find a polynomial $P_l(\cdot)$
  of low order (e.g., 9th order), such that the distribution of
  $P_l(|y_l(t)|)$ fits best with the desired distribution of the
  arguments (we use a
  Gaussian with mean $m=1$ and variance $\sigma^2=0.04$).%
  \footnote{$P_l(\cdot)$ is determined such that
    \begin{align}
      \label{gaussfit}
      \min=\sum_{s=0}^{N-1}
      \left[
        P_l\left(\left|y_l\left(Q_l(s)\right)\right|\right)-
        m-2^{1/2}\,\sigma\,{\erf}^{-1}
        \left(
          \frac{2 s+1}{N}-1
        \right)
      \right]^2,
    \end{align}
    where $Q_l(\cdot)$ is the permutation which rank orders the
    sequence $\{|y_l(t)|\}_t$ and ${\erf}^{-1}(\cdot)$ denotes the
    inverse error function given by
    \begin{align}
      \label{def_inverf}
      x=\frac{2}{\sqrt{\pi}}\int_0^{{\erf}^{-1}(x)}e^{-u^2}du \quad
      \text{for all $-1 < x < 1$.}
    \end{align}
%    However, the details of the fitting procedure are not crucial for
%    the algorithm to work.
    }$^,$%
  \footnote{\citet{schreiber96:_improv_surrog_data_nonlin_tests}
    achieve a similar result by using, instead of $P_l(\cdot)$,
    functions $M(\cdot)$ which map the sets of values of the time
    series (e.g.,$\{|y_1(t)|\}_t$) onto a given target set of values
    in such a way that rank ordering is preserved.  The method used
    here has the advantage over their method that the
    periodograms can be matched to arbitrary accuracy.}  %
  Construct two new complex-valued time series by using the arguments
  $P_l(|y_l(t)|)$ and the phases from step \ref{newphases}, and
  assign them to $y_1(t)$ and $y_2(t)$.
\item Repeat steps \ref{startloop}--\ref{endloop} until the
  periodograms of $y_1(t)$ and $y_1(t)$ converge to the same values,
  i.e., $\delta:=N^{-1}\sum_\omega (|\tilde y_1(\omega)|-|\tilde
  y_2(\omega)|)^2 < \epsilon$, with some small $\epsilon>0$ (e.g.,
  $\epsilon=10^{-6}$).  For larger $N$ convergence is substantially
  improved by judiciously choosing new permutations $\Pi_l$ from time
  to time (see below).
\item Multiply random phases $\exp(i\phi_l)$ to $y_1(t)$ and $y_2(t)$
  in order to make the distributions of $x_1(t)$ and $x_2(t)$ obtained
  in the next step time-translation invariant.
\item \label{makereal} Calculate the real-valued time series
  $x_l(t)=\mathrm{Re}\{\exp(i\omega_0 t) y_l(t)\}$ with some not too
  small $\omega_0$ (e.g., $\omega_0=5\pi/32$).
\item As in steps \ref{startloop} and \ref{retransform}, remove the
  small differences in the periodograms of $x_1(t)$ and $x_2(t)$
  resulting from the $\mathrm{Re}\{\cdot\}$ operation in step
  \ref{makereal}. But now, do it in a single ($100\%$) step.
\end{enumerate}
A typical result of running this algorithm is shown in
Fig.~\ref{fig:random}.  With a high enough frequency of the fast
modulation $\omega_0$, the algorithm guarantees that the origin of
delay space [$\Delta t\approx\pi/(2\omega_0)$] is encircled at finite
distance (see Fig.~\ref{fig:phasespace}).

\begin{figure}[h]
  \centering
  \epsfig{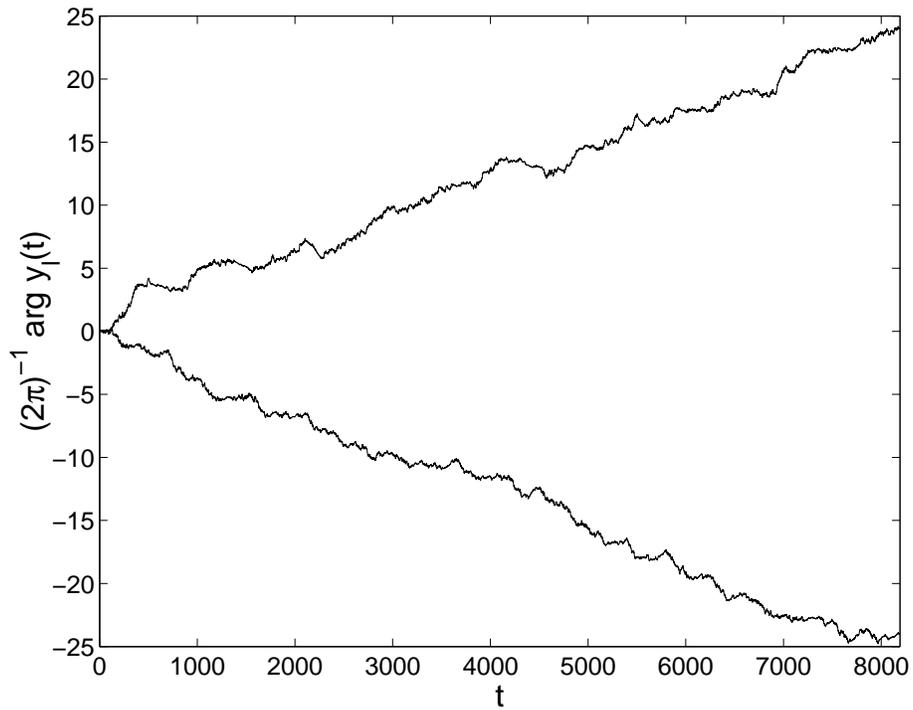}\\
  \caption{The phase angles of the complex time series $y_1(t)$
    (lower) and $y_2(t)$ (upper) that were used to generate $x_1(t)$
    and $x_2(t)$ of \emph{Pair B} (see text).  The two time series
    differ by $48$ cycles, and so do the time series $x_1(t)$ and
    $x_2(t)$.  Notice that $y_1(t)\times \exp(i\omega_0 t)$ and
    $y_2(t)\times \exp(i\omega_0 t)$ can be approximated by the
    analytic signals corresponding to $x_1(t)$ and $x_2(t)$.}
  \label{fig:phase}
\end{figure}

Now, some details concerning the manipulation of phases
(Steps~\ref{start_other}-\ref{newphases}) shall be explained.  The
permutations of the phase increments prior to the relaxation of large
phase jumps in steps \ref{diffusion} and \ref{diffusion_loop} prevent
the algorithm from running into solutions with localized regions of
fast drift of phase, which do not occur in typical experimental time
series.  However, it turns out that, with permutations selected with
equal probability from the set of all permutations, the algorithm has
difficulties to converge for large $N$, while without any permutation
it readily does.  This dilemma can be solved by slowly interpolating
from fully random permutations to the identity permutation: Execution
of the main loop (steps \ref{startloop}-\ref{endloop}) is dived into
blocks of, e.g., 200 iterations.  After terminating a block, the
$y_l(t)$ are reset to the pair of series with the lowest $\delta$ so
far, and the random permutations $\Pi_l$ are replace by the
permutations which rank order the sequences $\{(t+h
\nu_l(t))\mathop{\mathrm{mod}} N\}_t$ ($t=0,\ldots,N-1$).  Here
$\{\nu_l(t)\}_t$ are sequences of independent random numbers
distributed equally between $0$ and $1$ which are renewed after each
block and, as usual, $u \mathop{\mathrm{mod}} N:=u-m\,N$, where $m$ is
the largest integer with $u-m\,N\ge 0$.  The parameter $h$ controls
the interpolation between fully random permutations ($h\to\infty$) and
the identity ($h=0$).  Successively smaller values of $h$ are used
after each block.  As an example for a pair of long time series
($N=8192$, $-k_1=k_2=24$), \textit{Pair~B} was generated using
$h=400,200,100,0,0$ (convergence is reached after 5 blocks) and all
other parameters as for \textit{Pair~A}.  For these long time series
the fluctuations of the phases are clearly superimposed by a linear
drift of the phase difference (Fig.~\ref{fig:phase}).  For
\emph{Pair~A} this is not obvious, and one might be tempted to
attribute the difference in frequency there to a ``mere fluctuation''.
A Matlab code that implements the algorithm in the form used to
generate \textit{Pair~B} is available on the
internet\footnote{\url{http://www.fdm.uni-freiburg.de/$\sim$axel/onthelimits/}}.

\section{The influence of end-point effects for two spectral estimators}
\label{sec:estimators}

Calculating the periodogram of a time series (and the empirical
autocorrelation function as the Fourier transform thereof) is the
correct first step toward estimating its power spectrum only if
end-point effects can be neglected.  This is the case when the time
series allows a meaningful periodic continuation (which the algorithm
presented here guarantees), or if the time series are sufficiently
long.
For two methods of spectral estimation it shall here be demonstrated
that the condition of sufficient length can also be met by the
generated pairs of time series -- without making used of periodic
continuation.  That is, although the two methods differ in the way
endpoints are treated, in the way the periodograms are averaged, and
also slightly in their results, each of them gives virtually identical
results for the two time series of a pair.

\begin{figure}[h]
  \centering
  \epsfig{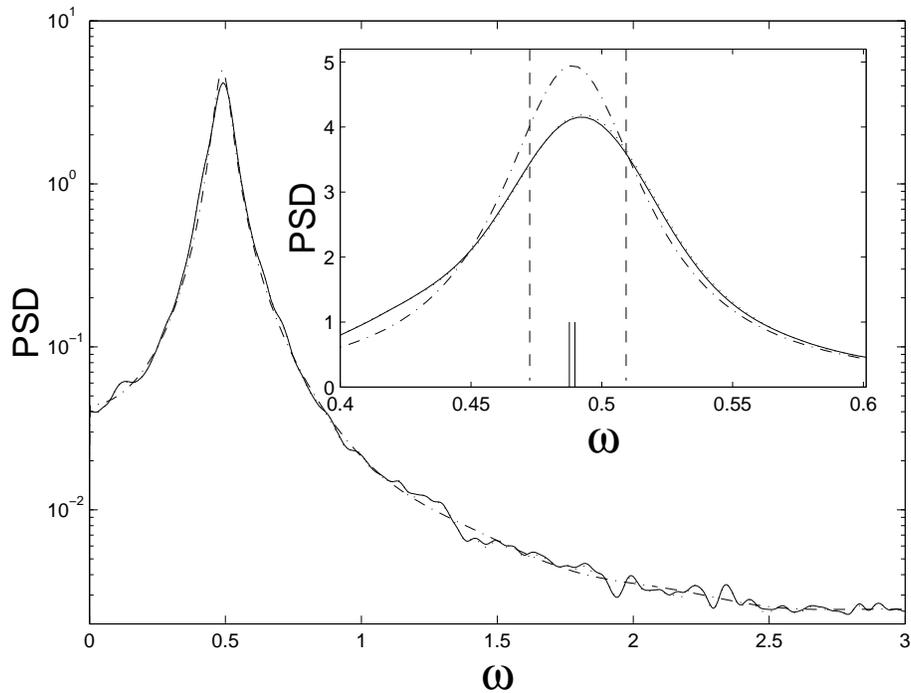}\\
  \caption{Estimated power-spectral densities (PSD, in units of power per
    radians per sample) for \textit{Pair~B}.  The two estimates
    (dash-dotted) using autoregressive modeling (see text) are
    indistinguishable.  When using Welch's Method (see text), the
    differences between the estimates corresponding to $x_1(t)$
    (solid) and $x_2(t)$ (dotted) are hardly visible.  The blowup of
    the peak region in the inset also shows the oscillation
    frequencies of $x_1(t)$ and $x_2(t)$ (vertical dashed lines) and,
    as an estimate of the nominal error of the autoregressive spectral
    estimates, the interval covered by the maxima of the spectra
    estimated by using only half the data of each time series
    (vertical solid lines near $\omega$ axis).  Notice that the inset
    has all-linear axes, while the large graph is semi-logarithmic.}
  \label{fig:power}
\end{figure}

First consider Welch's method
\citep{welch67:_use_fast_fourier_trans_estim_power_spect}.  Both time
series of \textit{Pair B} are split into $126$ segments of $192$
samples length with $2/3$ overlap.  The segments are tapered using a
Hamming window and, in order to interpolate the resulting power
spectra, extended to length 2048 by appending zeros.  The periodograms
of all segments of a time series are averaged.  As is shown in
Fig.~\ref{fig:power}, the estimated power spectra are nearly identical
for both series.  The differences are much smaller than then the
difference of the oscillation frequencies.

Next, autoregressive spectral estimation is used.  Applying Burg's
method \citep{marple87:_digit_spect_analy}, a $p$-th order
autoregressive process
\begin{align}
  \label{autoregressive}
  x(t)=\left(\sum_{k=1}^{p} a_k x(t-k)\right) + e(t)
\end{align}
with uncorrelated, identically distributed $e(t)$, is fitted to the
two time series of \textit{Pair~B}.  The resulting estimates for the
coefficients $(a_1,...,a_p)$ for $p=10$ are listed in
Table~\ref{tab:coefficients}.  No attempt was made to find the
``optimal'' model order $p$, since for all reasonable values of $p$
the same observation is made: The power spectra of the two fitted
autoregressive processes are virtually indistinguishable.  For
example, with $p=10$ the spectral maxima are both located at the
angular frequency $\omega:=2\pi f=0.4887$.  They differ only by
$\Delta\omega=2\times 10^{-6}$.  In order to get an estimate for the
errors of the autoregressive spectral estimates, the first and second
half of the two time series are fitted separately.  Then the enforced
property of having equal periodograms is lost and the estimated
spectra differ slightly.  The maxima of the resulting power spectra
are located at $\omega=0.4876$ and $0.4897$ for $x_1(t)$, and $0.4890$
and $0.4884$ for $x_2(t)$.  For comparison, the oscillation
frequencies of $x_1(t)$ and $x_2(t)$ are $\omega=0.4725$ and $0.5093$
respectively (see inset of Fig.~\ref{fig:power}).  Hence, it is not
possible to explain the undetectability of the difference in frequency
between $x_1(t)$ and $x_2(t)$ of \textit{Pair~B} by a lack of spectral
resolution.

\section{Conclusion}
\label{sec:conclusion}

Several examples were given that demonstrate that the power spectra of
time series alone do not contain sufficient information to ensure that
the oscillation frequencies (cycles per unit time) of the time series
could precisely be determined.  These examples illustrate a
corresponding general theorem \citep{rossberg02:_power_not_frequency}.
Furthermore, it was shown that this result does not depend on the
particular method of spectral estimation, as long as it is functional
in the periodogram or the empirical autocorrelation function,
respectively.
There is also no other method of evaluating the periodogram which
would yield the oscillation frequency.  This is shown by introducing a
general iterative method to generate pairs of time series of arbitrary
length which have identical periodograms but different frequencies of
oscillation.

  \newpage

%\bibliographystyle{ijbc}
%\bibliography{/home/axel/bib/bibview}

\vspace*{\fill}

\newpage 
\begin{table}[p]
  \centering
  \begin{tabular}{|l|c|c|c|c|c|}
    \hline
    &$a_1$&$a_2$&$a_3$&$a_4$&$a_5$\\
    \hline
    $x_1(t)$&1.1631 &-0.3064 &-0.1892 &-0.0787 &-0.0563 \\
    \hline
    $x_2(t)$&1.1632 &-0.3065 &-0.1892 &-0.0788 &-0.0558 \\
    \hline
    \hline
    &$a_6$&$a_7$&$a_8$&$a_9$&$a_{10}$\\
    \hline
    $x_1(t)$&-0.0198 &-0.0304 &0.0006 &-0.0072 &-0.0200\\
    \hline
    $x_2(t)$&-0.0198 &-0.0311 &0.0008 &-0.0067 &-0.0203\\
    \hline
  \end{tabular}
  \caption{The autoregressive coefficients in
    Eq.~(\ref{autoregressive}) with $p=10$ 
    estimated for \textit{Pair~B} (see text)
    using Burg's method.  The variance of $e(t)$ is estimated as
    $\mathop{\mathrm{var}}(e)=0.0409$ for both time series.}
  \label{tab:coefficients}
\end{table}

\clearpage

\section*{Figure Captions}

\vspace*{\fill}
\vspace*{\fill}

\end{document}